\def\year{2018}\relax
\documentclass[letterpaper]{article} 
\usepackage{aaai18}  
\usepackage{times}  
\usepackage{helvet}  
\usepackage{courier}  
\usepackage{url}  
\usepackage{graphicx}  
\usepackage{hyperref}
\graphicspath{ {./Images/} }

\begin{document}
%
\title{Intelligent Subset Selection of Power Generators for Economic Dispatch}
\author{Biswarup Bhattacharya\\
University of Southern California\\
Los Angeles, CA 90089. USA.\\
Email: bbhattac@usc.edu\\
\And
Abhishek Sinha\\
Adobe Systems Incorporated\\
Noida, UP 201301. India.\\
Email: abhishek.sinha94@gmail.com\\
}
\maketitle
\begin{abstract}
Sustainable and economical generation of electrical power is an essential and mandatory component of infrastructure in today's world. Optimal generation (generator subset selection) of power requires a careful evaluation of various factors like type of source, generation, transmission \& storage capacities, congestion among others which makes this a difficult task. We created a grid to simulate various conditions including stimuli like generator supply, weather and load demand using Siemens PSS/E software and this data is trained using deep learning methods and subsequently tested. The results are highly encouraging. As per our knowledge, this is the first paper to propose a working and scalable deep learning model for this problem.
\end{abstract}

\section{Introduction}
Power grids are one of the most vital components of infrastructure in modern society. The main objective of power grids is to provide continuous power to consumers (both household and industrial). In today's world, technology has taken the forefront in automating almost all aspects of life. With respect to electrical power grids, all units starting from generation to distribution have been installed with ``smart meters", ``smart" appliances to make the grid ``smarter". By definition, a smart grid is an electrical grid which includes a variety of operational and energy measures including smart meters, smart appliances, renewable energy resources, and energy efficient resources. Smart grid does not only mean hardware changes but also algorithmic changes. 

With increasing technological advancement and efficiency, the demand for power has increased to the point that resources need to be conserved carefully to generate power in optimal amounts so that no power is wasted and there is no shortage. This is an especially difficult problem to solve given the dynamic market situations where there are multiple parties involved, including businesses and governments. Also, for this to work, businesses should be making profits all the while generating and distributing enough power and the government's job is to regulate appropriately so that no region is left disadvantaged.  To enhance data visualization schemes, synchrophasors projects have recently been deployed with the help of which, the magnitude and angle of each phase of the three phase voltage and/or current, frequency, rate of change of frequency and angular separation at every few millisecond interval (say 40 milliseconds) can be monitored \cite{synchrophasor}. To the question of optimal generation, there are various factors like weather (solar energy), renewable/non-renewable source, generation capacity, transmission capacity, storage capacity, congestion, fault analysis \cite{bhattacharya1} among others involved which makes optimal power generation and distribution an even more daunting task. Thus, with the appropriate infrastructure, this paper attempts to design and explore a ``smart" algorithm which can efficiently select generator sources given the expected load demand.

Electric load is highly dependent on certain weather variables like temperature, humidity, and wind speed. Historical load data and weather information are used by utilities to generate short (up to a week), medium (one week to a year) and long-term (more than a year) load forecasts. As a result, so far, there has only been variability and uncertainty on the demand side of electric power given that majority of demand is supplied by fossil fuel and nuclear-based electricity in most of the world. Variability and uncertainty on both the supply and demand side will require a major rethinking of how to manage the electricity grids most optimally from the points of view of capacity planning, dispatch and overall risk management in order to meet the electricity demands with the expected reliability in the most economically efficient way.

\section{Related Work}
Majority of the literature that exists on subset selection is usually focused on portfolio optimization and statistical techniques, with heavy inspirations from financial models and asset management. The methodology adopted by us using state-of-the-art AI techniques has not been attempted before.

Some of the existing literature on subset selection and economic dispatch are discussed here. In \cite{rel1}, they have created a framework for efficiently managing the weather-related uncertainty risk of solar generators and facilitating their integration into power grids while optimizing the economic return from solar investments. They have used general optimization techniques. In \cite{rel2}, they proposed an innovative framework for analyzing the renewable generators at a given location and constructing energy portfolios that minimize the variability and forecasting error of the overall power output. In \cite{rel3}, they developed a model for analyzing the system imbalance with different mixes of renewable generation (solar photovoltaic and wind) including in the presence of large pools of plug-in electric vehicles (PEVs) that participate in vehicle-to-grid and grid-to-vehicle operations. This demonstrated how to mix sources to create a diverse portfolio. Economic dispatch (ED) is looked at with the lens of particle swarm optimization (PSO) and genetic algorithms. In \cite{rel4}, the paper presents a new genetic approach for solving the economic dispatch problem in large-scale power systems and can even take network losses, ramp rate limits, and prohibited zone avoidance into account because of genetic algorithm's flexibility. In \cite{rel5}, the proposed PSO method is demonstrated for three different systems, and it is compared with the GA method in terms of the solution quality and computation efficiency. In \cite{rel6}, a modified PSO (MPSO) mechanism is suggested to deal with the equality and inequality constraints in the ED problems. Our approach is quite different compared to these existing works as we have used a machine learning and deep learning approach to choose subsets. We have also taken inspiration from \cite{rel7}, where a dynamic model of the wholesale energy market due to the network constraints is derived to further understand the energy market especially in relation to smart grids.

\section{Problem Description}
Due to the weather-dependent nature of a number of renewable energy sources including solar and wind, uncertainty in relevant weather variables results in uncertainty in the power output. In order to participate in the electricity grids with a day-ahead commitment \cite{3almeshaiei}, power generators may have to guarantee a reliability at par with the traditional (fossil-fuel based) generators like thermal, which may mean 90\% or higher probability that the power supply commitment would be met. 

Similarly, in remuneration schemes where renewable generators have to pay for the power imbalance between their contracted amount (typically in the day-ahead market) and the actual output at operating time, power forecasting risk translates into risk to income and profitability of the renewable power suppliers. A technique that can reduce this power forecasting uncertainty for renewables will have a significant impact in maximizing the contribution to the grid from weather-dependent renewable generators as well as extracting higher economic returns from investments in renewable energy. The generation also needs to be done in a way that the power delivered by the renewable sources meet the load demand and the power generated does not cause congestion in the grid. 

Thus given a grid consisting of \textit{m} locations whereby the renewable power generators are present, our objective boils down to selecting a subset \textit{k} of these generators such the following three constraints are met:
\begin{itemize}
\item Economic return from generation is maximized.
\item Power generated is not lesser than the load demand.
\item Power generation does not induce congestion in the network.
\end{itemize}

\section{Dataset}
\subsection{Software for Data Collection}
For our simulation and testing purposes, we generated data using simulation tools like Siemens PSS/E \cite{1siemens} and PowerWorld Simulator \cite{2powersim}.

we utilized the useful scripting system named {\tt{psspy}} inbuilt with PSS/E with the help of which one can create grids, simulate situations using Python. 

\subsection{Making the Grid}
For the purpose of generation of data, it was required that simulations be run on an electrical grid. This is to enable collection of electrical voltage and phase at every bus i.e. to capture the state of the grid. An important aspect of the experiments was dealing with congestion and hence it was important to make the grid by hand such that congestion scenarios could be created more effectively. A well connected grid usually compensates the latent congestion in the grid. Thus, a grid was created by us completely by hand. 

The network (electric grid) used by us consisted of $20$ buses with $6$ generators and $8$ loads at various buses throughout the network. The base frequency used was $50$ Hz which is the standard for India. The base MVA for the network was $100$ MVA. Out of the $6$ generators, three of the generators were renewable (solar) and three of the generators were coal-based. There were two major tie-lines which connected the generator and load cutsets. It was expected that majority of the congestion will happen on these two major tie-lines. There was a fair amount of power distribution done among the generators as well as the load to simulate actual scenarios.

\subsection{Solar Data}
For the purpose of this experiment, we required solar data across multiple days and multiple time instants over each day. Some sources explored to get this data include but is not limited to Weather.com \cite{weathercom}, AccuWeather \cite{accuweather} as well as various sources cited in various research papers. However, due to lack of availability of free to use, well formatted and/or desired data, finally, the data used was from the IIT Kharagpur Electrical Engineering department rooftop solar panels. Scaling of the data was done appropriately to align it to the grid designed. General trend of the data was seen and analyzed with the help of insights from \cite{8sharma}.

\subsection{Congestion}
Transmission congestion occurs when there is insufficient energy to meet the demands of all customers. No actual congestion occurs in the transmission system; these systems do not slow down, and electricity does not become blocked or delayed because the transmission system cannot be stretched beyond its limits. Attempting to operate a transmission system beyond its rated capacity is likely to result in line faults and electrical fires. Congestion happens when there is a shortage of transmission capacity to supply a waiting market. During congestion, systems run at full capacity and proper efficiency which cannot serve all waiting customers \cite{congestiondefn}.

In a competitive market, regulatory bodies are aware of the risk of price gouging from utilities that control transmission services due to congestion, and most have safeguards in place to insure that abusive pricing does not occur. The only ways the congestion can be alleviated are to tune the system to increase its capacity, add new transmission infrastructure, or decrease end-user demand for electricity.

\subsection{Methodology}
\subsubsection{Data Generation for estimating congestion}
For the simulation to get data for the module which checks if there is congestion or not, the data generation was fairly straightforward. For a given configuration of generators and loads in the grid, a load flow solution is run using {\tt{psspy}}. This simulation was run for Low, Medium as well as High Loading. As some of the generators were solar, so generation varied throughout the day. Thus, the grid experienced different amounts of total generation throughout the day. Hence, there is both differential loading as well as differential generation, leading to a lot of possible load flow states throughout the day. 

We took data for 16 days from March 1, 2017 to March 14, 2017 (from the Electrical Engineering Department solar cells). Thus, essentially there were $14$ days worth of data with around $51$ different levels of generation throughout the day. This led to a total of $714$ instances for the time period. On top of this, there were $3$ different types of load. So, totally, there were $714 \times 3 = 2142$ load flow solutions run to generate the required states. This state consisted of voltage and angle at all buses of the grid.

\subsubsection{Data Generation for actual subset selection}
For the simulation to get data for the module which decides which subset of generators is the most optimal choice, the data generation was much more involved. The main motivation behind generating data was to simulate the various scenarios which can happen due to only a subset of the generators being on.

There were $3$ solar generators and $3$ coal generators in the grid. Our objective is to find out switching on which of the generators will meet the demand at the least cost. To achieve this, it was essential to see how the load flow changed on switching off some of the generators. Thus, as the base, initially all 6 generators were kept switched on and then a load flow was run to get the state at $t=0^{+}$. Then, at $t=1$, a subset of the generators were switched off and the load state was run again. The following combinations of generators were switched off to make various situations to be simulated:
\begin{itemize}
    \item Only solar generator 1 is switched off
    \item Only solar generator 2 is switched off
    \item Only solar generator 3 is switched off
    \item Solar generators 1 \& 2 are switched off
    \item Solar generators 2 \& 3 are switched off
    \item Solar generators 1 \& 3 are switched off
    \item All solar generators (1, 2 \& 3) are switched off
\end{itemize}

The above combination was performed for all instants of the day over the $14$ day period as mentioned above. According to the solar data collected by us, the generation values were available at every 15 minute interval (e.g. starting from 12 midnight, next value at 12:15 AM, then 12:30 AM and so till 11:45 PM that day). Thus, if we took $t=0$ to be 12 AM then $t=1$ would be 12:15 AM. It must be noted that at $t=1$ while running the simulation, we used the power values the same as that of 12:15 AM if $t=0$ is 12 AM. Essentially we were seeing one step jumps across the day and how it would affect the grid if a combination of solar generators were switched off. Also, after every simulation, we checked if congestion was taking place using the previous architecture. If congestion was taking place after switching off some combination of generators, then clearly choosing that subset would be detrimental.

For every instant of the day we had $7$ possibilities (the above $7$ combinations) that can happen for the next time instant. Therefore, total $700 \times 7 = 4900$ simulations (some time instants like $7$ PM to $5$ AM were discarded as solar generation is almost zero and we assumed battery power was not available) were run for one type of loading. This data generation was repeated for Low, Medium and High loading.

\section{Predicting congestion in the grid}
\subsection{The Input}
The data corresponding to a network having $20$ different buses was obtained. Number of different simulations were run to obtain data corresponding to both congestion and non-congestion. The size of the data set was $715$ observations out of which $650$ was used for training and the other used for testing. The initial voltage value corresponding to each of the buses along with the actual power at the $3$ solar generators would be fed as input to the model and the model was expected to output as to whether this initial condition would lead to congestion or not.

\subsection{Classification using SVM}
In machine learning, support vector machines (SVMs) are a traditional form of supervised learning algorithm which learns to classify into two classes based on the previously seen examples \cite{svm}.

Out of the $715$ data examples available, $650$ were used for training while the rest were used for testing. The input to the SVM model was the voltage data of all buses \& actual solar power data. The output of the SVM classifier was whether congestion occurs or not.

\subsection{Classification using deep neural networks}
A neural network with $1$ hidden layer consisting of $100$ neurons was constructed to do the prediction task. The input layer was of size $23$ corresponding to each of the bus voltages and the solar power values whereas the output was of size $2$ corresponding to the probabilities that the congestion would occur or not. To introduce non-linearity in the model \textit{ReLU} activation was used after the hidden layer \cite{relu}. The model was trained using the \textit{Adam Optimizer} \cite{11kingma} to minimize the training loss. Cross entropy loss was used as a measure of the training loss.

\subsubsection{ReLU Activation \& Adam Optimizer}
In the context of artificial neural networks, the rectifier is an activation function defined as: $f(x)=\max(0,x)$. A unit employing the rectifier is called as a rectified linear unit (ReLU) \cite{stanford231n}. 

Adam is a stochastic gradient descent algorithm based on estimation of $1^{st}$ and $2^{nd}$-order moments. Adam takes three hyper-parameters: the learning rate, the decay rate of $1^{st}$-order moment, and the decay rate of $2^{nd}$-order moment \cite{11kingma}.

The inputs and outputs of the model are the same as that of the SVM Classifier method.

\subsection{Results}

Using SVM, we find that the model obtained an accuracy of 93-94\% on the test data set. Using neural networks, after training the model for around $500$ steps, the final test accuracy on the test set was observed to be around 97-98\% thus showing an improvement of around 5\% compared to the SVM model. This is expected as the neural network is able to capture hidden information regarding the congestion situation more effectively than a simple SVM classifier.

The training accuracy plot has been shown in Figure 1 and the test accuracy plot has been shown in Figure 2.
\begin{figure}[h]
\centering
\includegraphics[width=\linewidth, keepaspectratio]{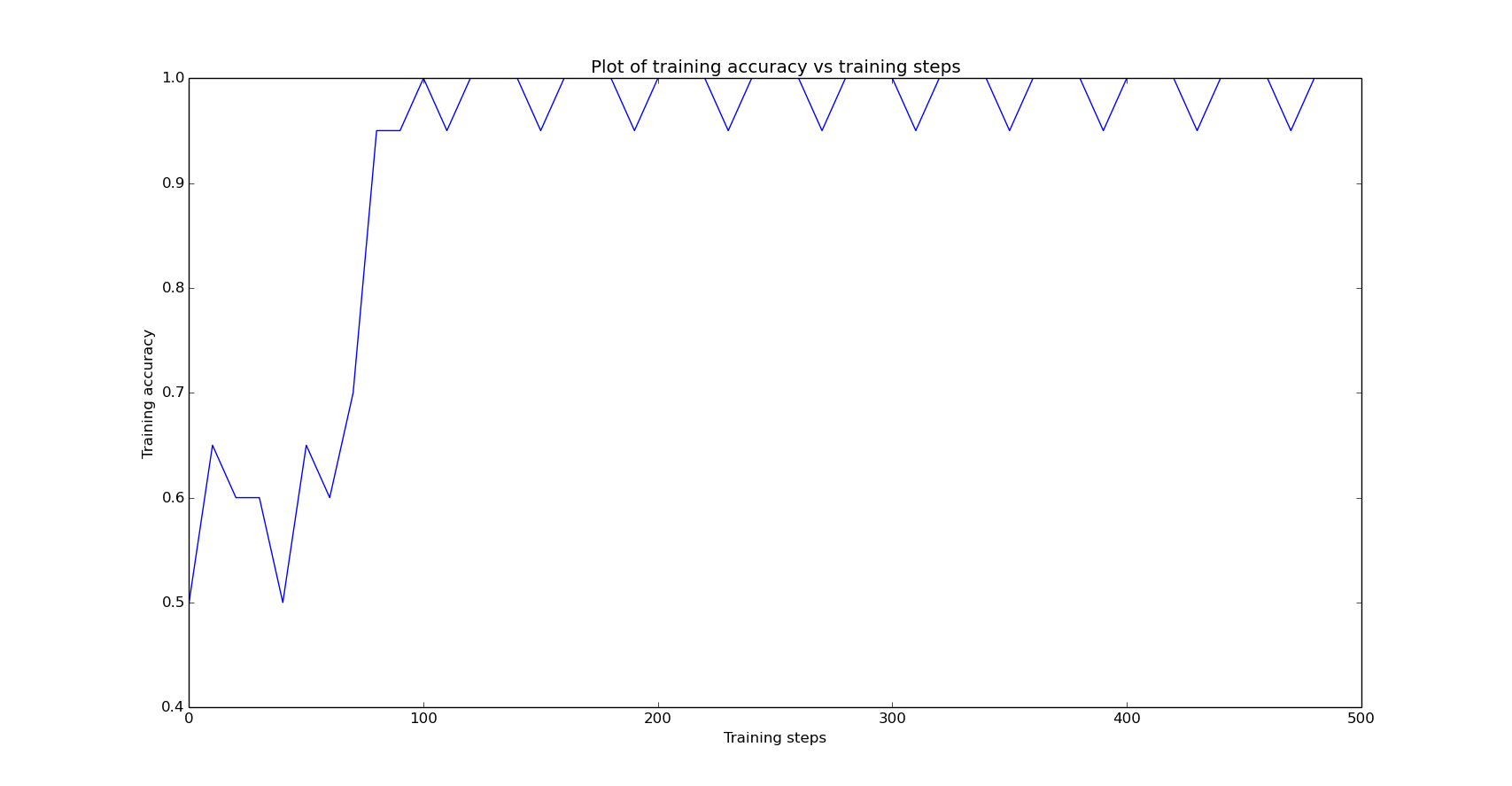}
\caption{Plot of training accuracy with training steps}
\end{figure}

\begin{figure}[h]
\centering
\includegraphics[width=\linewidth, keepaspectratio]{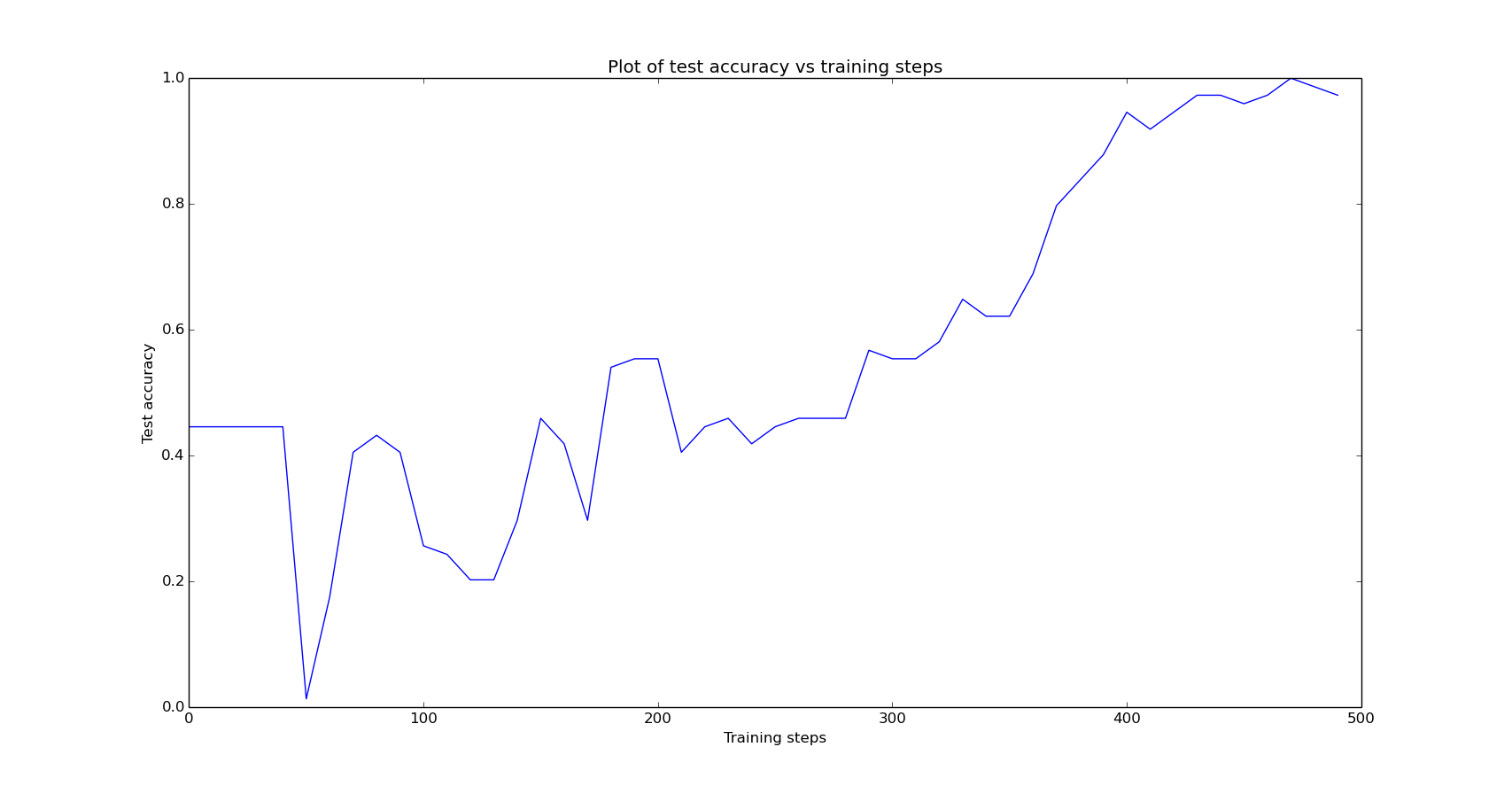}
\caption{Plot of test accuracy with training steps}
\end{figure}

\subsection{Prediction congestion when the actual solar power is not known}
In the previous part we assumed that we know the actual solar power from before. But in reality we know just the power committed by the solar power generators and not the actual power generated from before. So actually we just know the predicted load state of the network and the committed solar power from the generators. Now using the weather information along with the above knowledge we can predict if a congestion is going to occur or not in the network. As a substitute for both the committed power and the weather information we used the predicted solar power as input. The predicted power was calculated by averaging the past history of solar powers at different times of the day.

\subsubsection{The Input}
The model now again contains $23$ inputs, $20$ being the voltage values corresponding to the $20$ buses and the other $3$ the predicted solar power corresponding to the solar generators. The output of the model will be whether the input data would lead to congestion or not. $750$ data sets were generated again out of which $650$ were used for training and the rest for testing.

\subsubsection{Prediction results}
When SVM was used to classify the test data the testing accuracy was found to be 85\%. When the same neural network model described previously was used for classification, the test accuracy of the model was observed to be 91\% after 800 steps of training. This is expected as the solar power data was helpful in estimating the congestion situation, and not knowing it leads to uncertainty in the output.

The training loss of the neural network model is graphed in Figure 3 and the test accuracy has been plotted in Figure 4.

\begin{figure}[h]
\centering
\includegraphics[width=\linewidth, keepaspectratio]{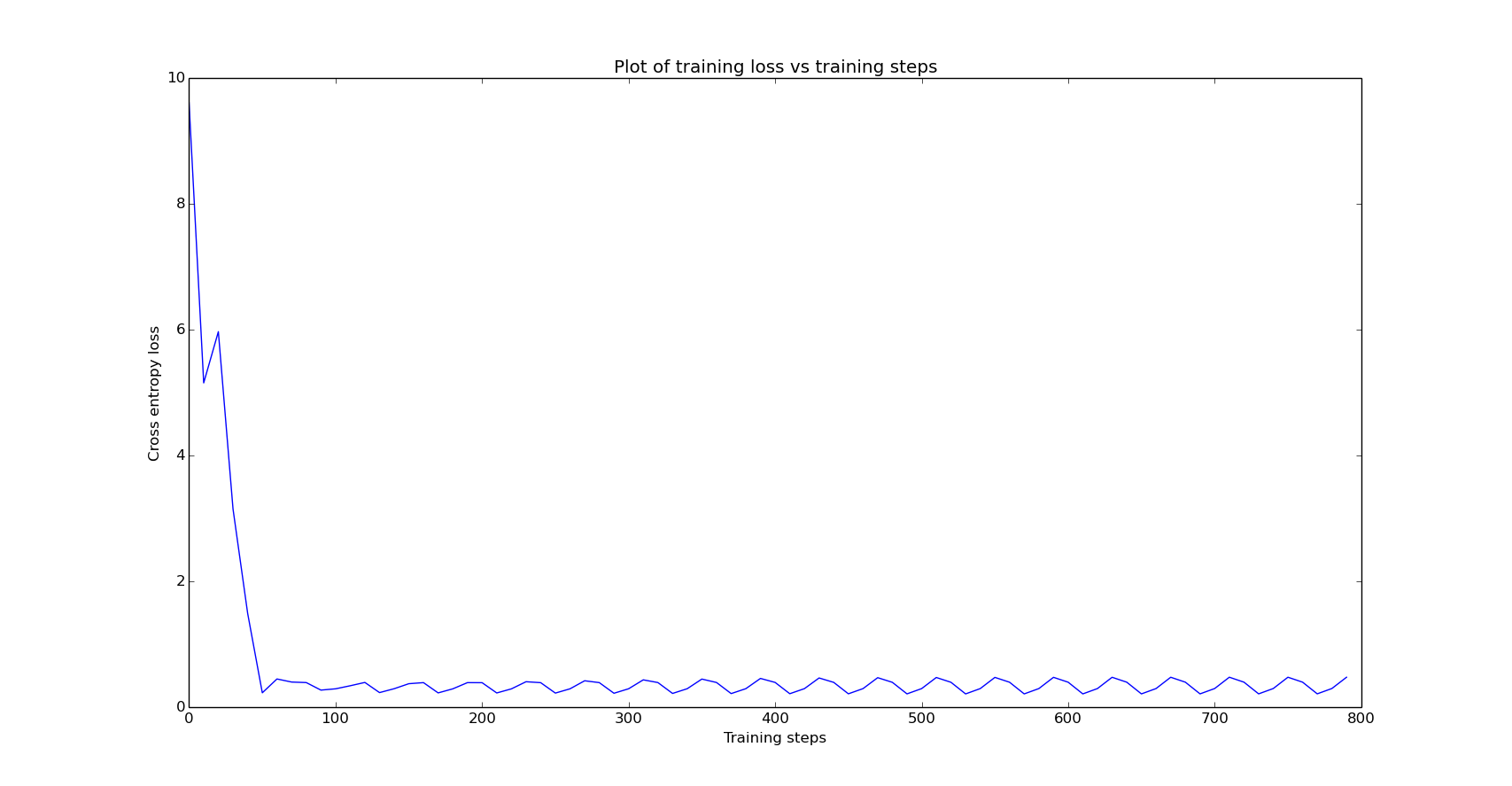}
\caption{Plot of training loss with progress in training}
\end{figure}

\begin{figure}[h]
\centering
\includegraphics[width=\linewidth, keepaspectratio]{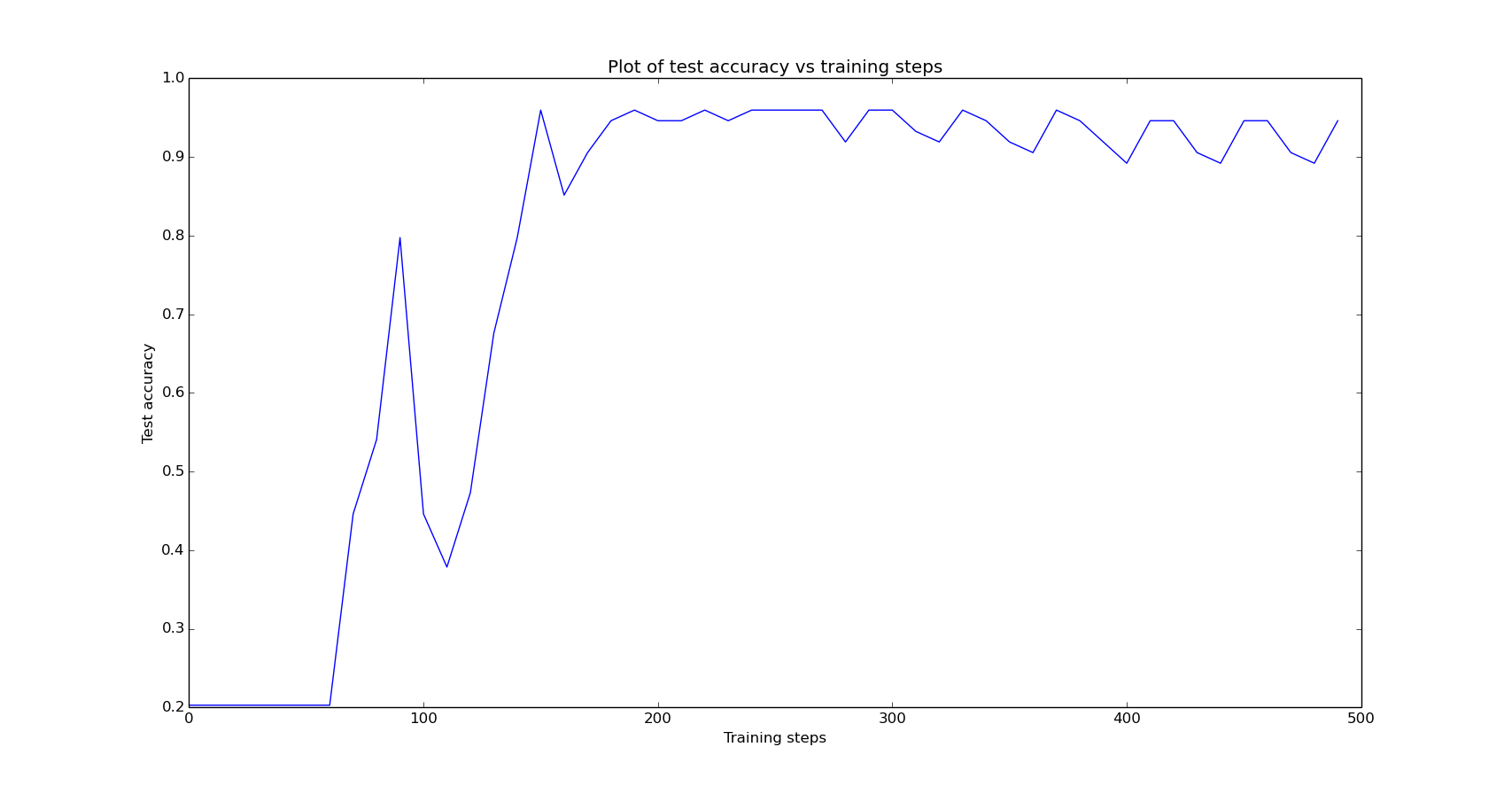}
\caption{Plot of test accuracy with progress in training}
\end{figure}

\section{Choosing the optimal subset of solar generators}
\subsection{Objective}
The task that we are trying to solve in this part is the  selection of the optimal subset of solar generators that minimize the total penalty occurred in using the solar generators. Given a past history of load state of a region and the solar power generated from the generators the model should be able to analyze the data in a way such that when given the predicted load state for the next day and the weather parameters for the day the model outputs the optimal subset of generators that minimize the penalty incurred.

\subsection{Penalty incurred corresponding to a subset}
Corresponding to any subset of generators, two types of penalties will be occurred by the network.

\subsubsection{Power mis-commitment penalty}
The first penalty that the generator operators have to face for a renewable source of generator is the mis-commitment penalty. Due to the uncertainty in the weather and thus the uncertainty in the power which will be generated, the power generators are often unable to generate the amount of power which they had promised to generate. The difference in the promised and the delivered power is covered up the non-renewable source generators but it comes up through an extra cost incurred by the grid operators. The larger is the difference between the committed and the delivered powers, the greater is the penalty incurred. Thus to choose an optimal subset one of our aim should be to choose the subset that shows lesser variance in the prediction depending on the current weather conditions.
Thus the loss $L_1$ used to train the network is given by:
\begin{equation} 
L_1 = (\textnormal{predicted power} - \textnormal{actual power generated})
\end{equation}

\subsubsection{Penalty incurred due to congestion in the network}
Even though the misprediction penalty corresponding to a subset can be lower than the other subset but the choosing the other subset might cause a congestion in the network leading to the further changes in the network to be done to resolve the congestion. Thus there comes a second penalty which is incurred if choosing a subset of solar generators leads to congestion in the network. The congestion loss $L_2$ is calculated as follows: if the subset causes a congestion in the network then $L_2 = 50$ otherwise $L_2 = 0$.

\subsection{The Model}
Simulations corresponding to a $20$ grid network consisting of 3 solar generators was run. For all the $6$ possible subsets of generators, the $L_1$ and $L_2$ values were computed. Loss $L_1$ is scaled accordingly so that the $L_1$ and $L_2$ values are approximately on the same scale. The input to the model would be the voltage corresponding to each of the $20$ buses before any of the subset has been chosen and the subset description. By subset description we basically mean information as to which generators amongst the $3$ solar generators are chosen and the predicted solar power at the chosen generators. Thus it can be described by a vector of size $3$ corresponding to the $3$ solar generators whereby an element of the vector will be zero if that generator was not chosen. If that generator was chosen then the vector corresponding to that index will contain the predicted power for that generator. Thus the input to our model is a vector of size $23$ $(20+3)$. The total loss corresponding to the subset is $L_1+L_2$.

The model would be expected to output the total loss incurred in selecting the input subset. Thus the loss function ($\mathcal{L}$) that the model tries to minimise is given by:
\begin{equation}
\mathcal{L} = (\textnormal{model output} - (L_1+L_2))^2
\end{equation}

A total of $4900$ data points were collected from the simulation out of which $4500$ were used as training data.

The model consisted of $1$ hidden layer with $200$ hidden neurons. The input layer is of size $23$ corresponding to the $23$ inputs whereas the output layer is of size $1$. To make the model complexer, ReLU activation was used after the hidden layers. Adam optimizer was used to train the model to minimize the loss $\mathcal{L}$.

\subsection{Results}
After training the model for around $2500$ steps the following results were obtained:
\begin{itemize}
\item Training minimum $L_2$ loss = $35$
\item Training minimum $L_1$ loss = $6$
\item Test $L_2$ loss = $59$
\item Test $L_1$ loss = $8$
\end{itemize}
These results are especially encouraging as the low test losses show the ability of our model to select an optimal subset which incurs the least penalty, i.e. most economical choice with lowest possible congestion probability.

The plots for the training and test $L_2$ loss have been shown in Figures 5 and 6 respectively.

\begin{figure}[h]
\centering
\includegraphics[width=\linewidth, keepaspectratio]{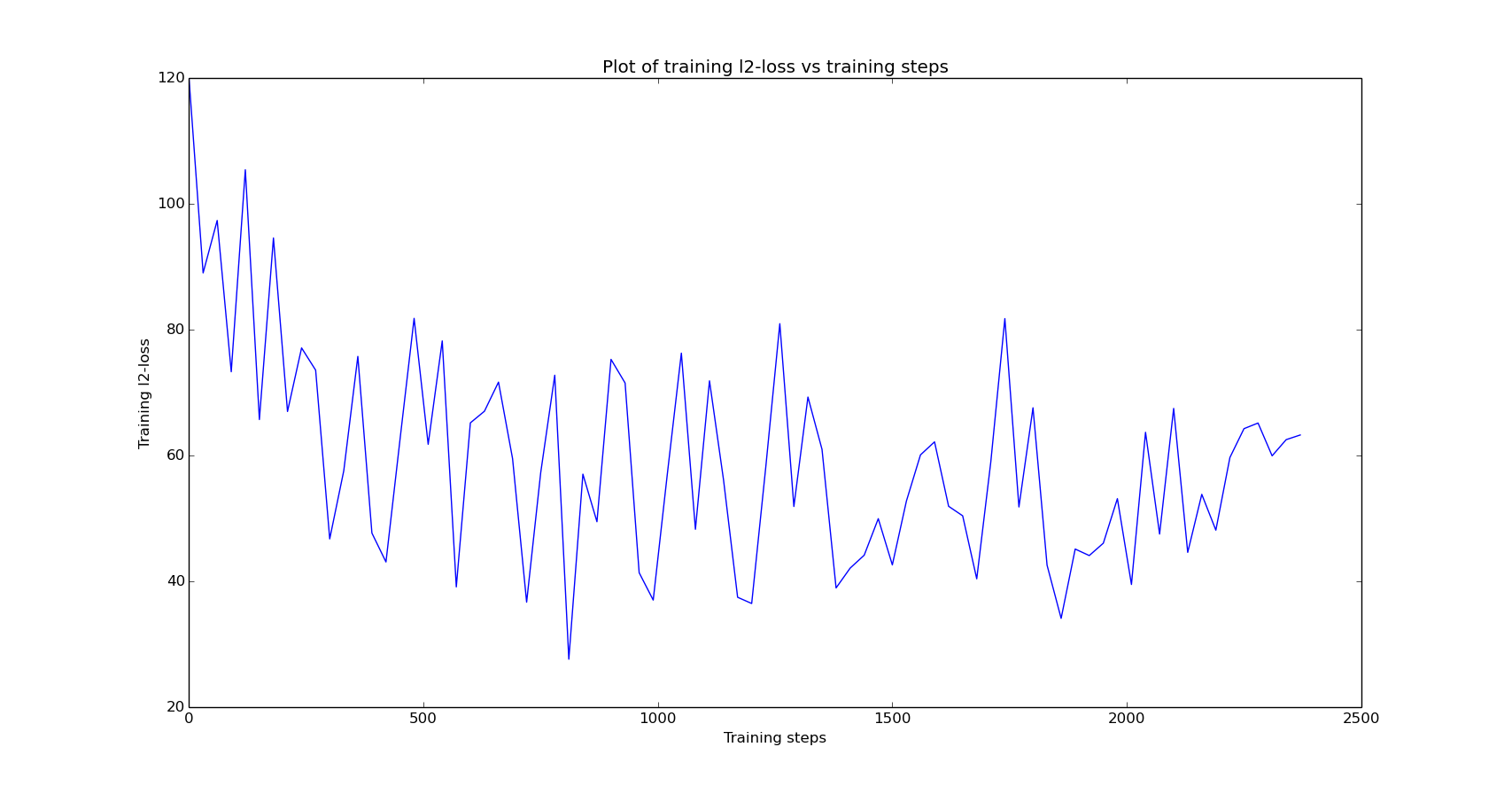}
\caption{Plot of training $L_2$ loss with progress in training}
\end{figure}

\begin{figure}[h]
\centering
\includegraphics[width=\linewidth, keepaspectratio]{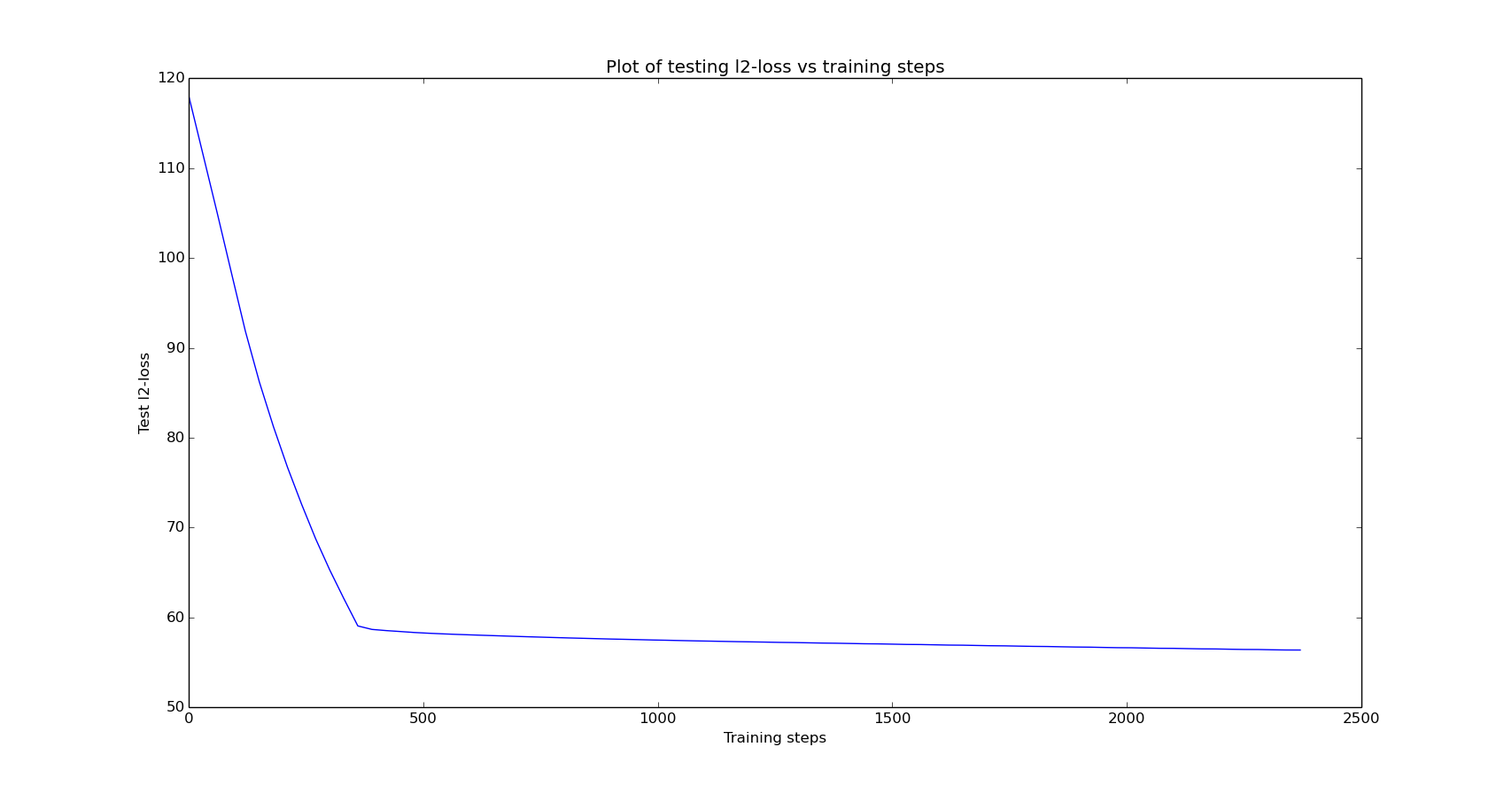}
\caption{Plot of test $L_2$ loss with progress in training}
\end{figure}

\section{Conclusion}
In conclusion, we created a grid to perform intelligent subset selection, by predicting congestion and selecting a economic choice of subsets having or not having solar generators, all using machine learning and deep learning techniques. The working system will be especially useful in renewable energy power grids where the generation periods are erratic (e.g. solar power generators generate power only during the day when the Sun is up). With the knowledge about the vulnerability of the grid, issues like load shedding, power surges etc. can be handled efficiently. The results obtained from our experiments were very encouraging and thus, further analysis can be done on various avenues starting from this. Some of these avenues are discussed henceforth. 

The advent of renewable electricity with its enormous potential and inherent regional and national character presents an opportunity to examine the local structure of the grid and establish coordinating principles that will not only enable effective renewable integration but also simplify and codify the grid's increasingly regional and national character. With time, the system will have the ability to be more sophisticated to handle various types of networks and situations which will be suitable for deployment at the national level.

\section{Acknowledgment}
The authors were undergraduate students of the Indian Institute of Technology (IIT) Kharagpur, West Bengal, India during the period of research for this project which culminated in this paper. They thank the institute for their support.

\fontsize{9.0pt}{10.0pt} 
\selectfont
\bibliographystyle{aaai}
\bibliography{biblio}

\end{document}